# Conversion of G-code programs for milling into STEP-NC

Shixin XÚ [1],[2], Nabil ANWER [1], Sylvain LAVERNHE [1]

(1) Laboratoire Universitaire de Recherche en Production Automatisée, ENS de Cachan, 94235 Cachan, France
{anwer; lavernhe; sxu}@ens-cachan.fr

(2) School of Mechanical Engineering and Automation, Beihang University, Beijing 100191, China
xushixin@buaa.edu.cn

**Abstract:** STEP-NC (ISO 14649) is becoming a promising standard to replace or supplement the conventional G-code programs based on ISO 6983 due to its feature based machine independent characteristics and its centric role to enable efficient CAD/CAM/CNC interoperability. The re-use of G-code programs is important for both manufacturing and capitalization of machining knowledge, nevertheless the conversion is a tedious task when carried out manually and machining knowledge is almost hidden in the low level G-code. Mapping G-code into STEP-NC should benefit from more expressiveness of the manufacturing feature-based characteristics of this new standard.

The work presented here proposes an overall method for G-code to STEP-NC conversion. First, G-code is converted into canonical machining functions, this can make the method more applicable and make subsequent processes easier to implement; then these functions are parsed to generate the neutral format of STEP-NC Part21 toolpath file, this turns G-code into object instances, and can facilitate company's usage of legacy programs; and finally, also optionally, machining features are extracted to generate Part21 CC2 (conformance class) file. The proposed extraction method employs geometric information of cutting area inferred from toolpaths and machining strategies, in addition to cutting tools' data and workpiece's dimension data. This comprehensive use of available data makes the extraction more accurate and reliable. The conversion method is holistic, and can be extended to process a wide range of G-code programs (e.g. turning or mill-turn codes) with as few user interventions as possible.

## 1- Introduction

Most CNC machines are programmed in the ISO 6983 G-code language, which limits program portability because the language focuses on coding the tool center path with respect to machine axes, rather than the machining process with regard to the part. Moreover CNC vendors usually extend the language beyond the limited scope of ISO 6983 creating their own macro-languages implying that they can only be executed on specific machine-tools. STEP-NC (STEP Data Model for Computerized Numerical Controllers) is a model of data transfer between CAD/CAM systems and CNC machines. It aims at standardizing the data formats used at the machine level, one key link in the entire process chain in a manufacturing enterprise. STEP-NC specifies machining processes rather than machine tool motion, using the object-oriented paradigm and the concept of "workingsteps", which correspond to high-level manufacturing features and associated process parameters. Thus STEP-NC creates an exchangeable, workpiece-oriented data model for CNC machine tools, supports the direct use of computer-generated product data from ISO 10303, and ensures compatibility of CNC input data. CNCs are responsible for translating workingsteps to axis motion and tool operation [I1].

In the course of STEP-NC adoption, there are many needs to convert legacy G-codes into STEP-NC programs. With the impending prevalence of this new standard, manual conversion of G-codes will be a huge, tedious task. Therefore, automatic and effective conversion will be highly adopted by manufacturing enterprises. On the other hand, legacy programs are important resources for enterprises. They contain optimal cutting conditions and machining strategies for making products, and embody implicitly the machining know-how from various experts. So converting legacy G-codes to get corresponding STEP-NC files, instead of designing from scratch, would save much costs and resources for enterprises. Also this practice would facilitate the accumulation of machining know-how due to information storage in object-oriented structures.

The essential of the conversion is reconstruction of a manufacturing feature oriented NC program from a G-code program. NC programming by feature approach can streamline the manufacturing cycle and make CAM/CNC integration easier since manufacturing feature data and the associated process data originate from CAM. Also the feature approach makes NC-to-CAM feedback link realizable. The core of STEP-NC data model is manufacturing features and machining operations, which are encapsulated in a workingstep. Figure 1 shows STEP-NC data model structure [I1]. A workingstep represents the machining process for a specified area of the workpiece. It specifies the association between "machining_feature" and "machining_operation" to be performed on the feature. The "machining_workingstep" is characterized by the use of a single tool and a set of technological parameters. A machining operation contains technological data for a workingstep. A workplan is a collection of workingsteps with an execution sequence. The "project" serves as a starting point for the program execution. The division of information means that changing the sequence of workingsteps or optimizing tool paths can be done with minimal impact on the rest of the data.

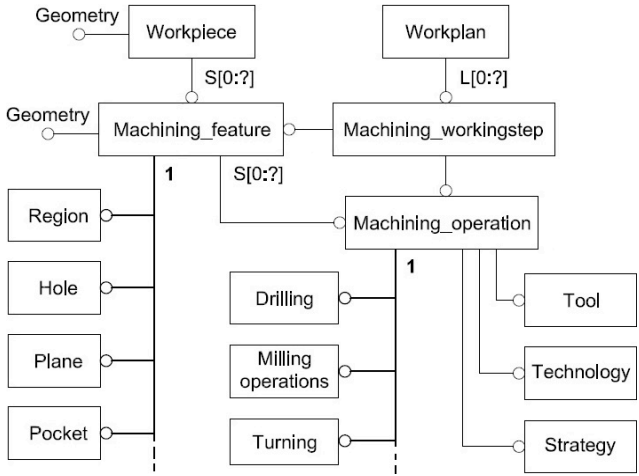

**Figure 1: STEP-NC data model.**

Only few researches addressed G-code to STEP-NC conversion. Shin et al. developed a system, called G2STEP, to convert G-codes into STEP-NC for turning applications [SS1]. The system is based on two main functions: the pre-processor function and the turning feature recognition function. In the pre-processor, each block of a G-code program is interpreted and stored in a pre-defined data structure. These blocks are divided using some hints and these blocks are grouped into workingsteps. At the turning feature recognition level, the feature profile remaining after a workingstep is generated, and the manufacturing feature, defined in STEP-NC, is recognized by a profile and pre-determined machining operation. However, this work did not handle milling applications and not fit well for roughing operations in G-code programs.

Zhang et al. proposed a method to re-use the process knowledge embedded in G-code part programs with different manufacturing resources [ZN1]. The authors emphasized the "process comprehension", which is "essentially restructuring the combined manufacturing information in an NC program into a high-level process plan and the associated resource information." They proposed an abstract meta-model for different CNC controllers. The method briefly involves how to get a STEP-NC file from a G-code. They recognized features from a G-code mainly by tool types, toolpath boundary and the rawpiece geometry. STEP-NC entities are then created. However details of feature types and feature parameters, are not mentioned in the paper. Their method tries to decode a G-code to get a STEP-NC file directly. Thus increasing the implementation difficulty and limiting the applicable scopes.

This paper aims to present a systematic method for automatically converting G-code part programs into STEP-NC formats. The method is applicable to both milling and turning G-codes. In Section 2, we first make clear what should be given as inputs to the conversion process, followed by the overall strategy of the conversion. In Section 3, detailed descriptions of the conversion of G-codes into canonical machining functions are described. In Sections 4 and 5, detailed descriptions for obtaining STEP-NC Part21 files, implemented in explicit tool path level, and in manufacturing feature level are presented. The system development and a testing example are given in Section 6, followed by concluding remarks in Section 7.

## 2- Method overview

Since there is no information about tools, rawpiece and its setup in a G-code, we should supplement them for ensuring successful conversion. And the G-code should be an error-free part program, hopefully, with high-quality, to make the conversion significant.

For example, one simple G-code with different shape of rawpieces will produce different final workpieces (Figure 2). With the following G-code, if a small rawpiece is used, it will make a profile feature; if a large rawpiece, it will make a slot feature. Similarly, the rawpiece's setup location, or its offsets with regard to the programming frame also affects the workpiece's geometry. Therefore the rawpiece's geometry, location, etc. are necessary supplements for the conversion. In fact, these supplements for the conversion roughly correspond to the phase of machining definition for carrying out a CNC machining.

```
G90 G54 G0 X0 Y0 Z100
Y-50
G1 Z-50 F200 S300 M3
X100
Y50
X-100
Y-50
X0
Z100
G0 Y0 M5
M30
```

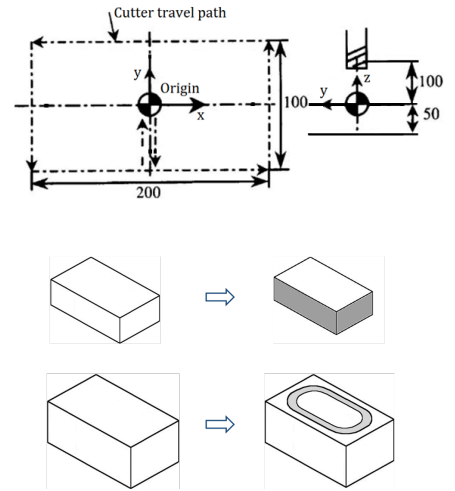

**Figure 2: From rawpiece to final workpiece.**

In this work, we consider G-code program, information of cutting tools used in the G-code, rawpiece's shape and location. Besides, we should know whether the given G-code is a milling program, or a turning one, as well as the number of axis involved in the program as far as current realization.

Two types of conversion outcomes are proposed: STEP-NC Part21 file in explicit toolpaths (conformance class 1, or CC1) and STEP-NC Part21 file in manufacturing features (CC2). The first type mainly use toolpath features, and can deal with 3 to 5-axis part programs. The second type performs manufacturing feature extraction from the given G-code part program, and it can deal with G-code programs for 2½D machining at present realization. The information flow is illustrated in Figure 3.

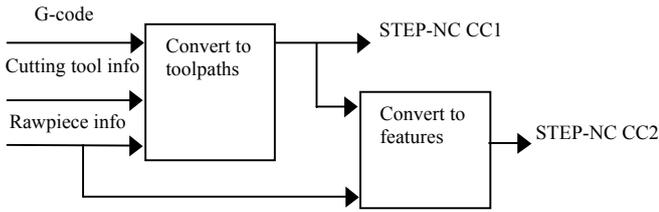

**Figure 3: The information flow.**

In order to illustrate the method, we use a very simple example as shown in Figure 4a. The case study is a finish milling in one layer to remove a 5mm-depth material on the planar top face of the stock. In this case the only machining feature is the planar face. The toolpath for one layer cutting is shown in Figure 4b, followed by the G-code in Table 1.

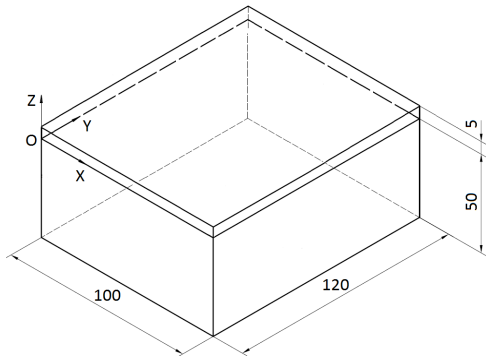

a) The rawpiece

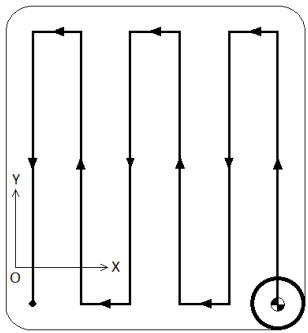

b) The toolpath

**Figure 4: Illustration example.**

**Table 1: G-code for milling the planar face**

| | |
|---|---|
| G54 G90 G21 G40 G49 M5 M9 | Y133.50 |
| T1 M6 (Use an endmill, Ø18mm) | X40.60 |
| G43 H1 (Cutter length compensation by 50mm) | Y-13.50 |
| M8 S720 M3 | X23.50 |
| G0 X91.90 Y-13.50 Z100.00 | Y133.50 |
| Z15.00 | X6.40 |
| G1 Z0.00 F240.00 | Y-13.50 |
| Y133.50 | G0 Z15.00 |
| X74.80 | G49 M9 M5 |
| Y-13.50 | M30 |
| X57.70 | |

The functions of any CNC machine can be viewed as a set of canonical machining functions defined by NIST based on ISO 6983 [K1]. If a machine has new functions beyond this standard, one can add new canonical functions. The G-code program is firstly mapped into canonical functions that the machine should execute. Secondly, by analyzing hints (such as tool changes, speed changes) in the canonical functions, workingsteps can be generated. In these workingsteps all operations are treated as freeform operations and features as toolpath features, except those that can be easily attached to operations, like canned cycles. Then toolpath data is converted into the data structure "toolpath" of freeform operations. Thus we can generate STEP-NC Part21 file in explicit toolpaths.

The "toolpath_feature" is introduced in STEP-NC to enable the definition of tool movements not covered by regular machining features (such as pockets, holes, slots, steps). It is a placeholder for explicit toolpaths assigned to the operations associated with it. 3 to 5-axis milling of freeform surfaces typically requires explicit specification of toolpaths. This kind of Part21 file (CC1) still has advantage over G-code. By connecting this information with the high-level operation and feature data, the toolpaths can always be interpreted within their semantic context. They are also provided in a structure which allows to identify the individual toolpath rather than to search through thousands of lines of unstructured code for axis movements.

Finally, we can extract manufacturing features from the toolpath data of the Part21 file by analyzing the machining regions, machining strategies, etc. In this process we need to merge some workingsteps and reorganize them. When achieved, the STEP-NC Part21 file in features and operations can be generated. The final step can be optional according to the needs of each company.

The overall procedures can be summarized as follows. A G-code program is firstly translated into canonical functions, which are then interpreted into a STEP-NC Part21 file (CC1). The CC1 file is further converted into the higher level STEP-NC Part21 file (CC2) after manufacturing feature extraction [I2] [I3] [I4]. If we want to share a G-code program in a higher level, or want to establish a bidirectional information flow between CNC and CAM, we can choose CC2 conversion. Figure 5 shows the overall strategy.

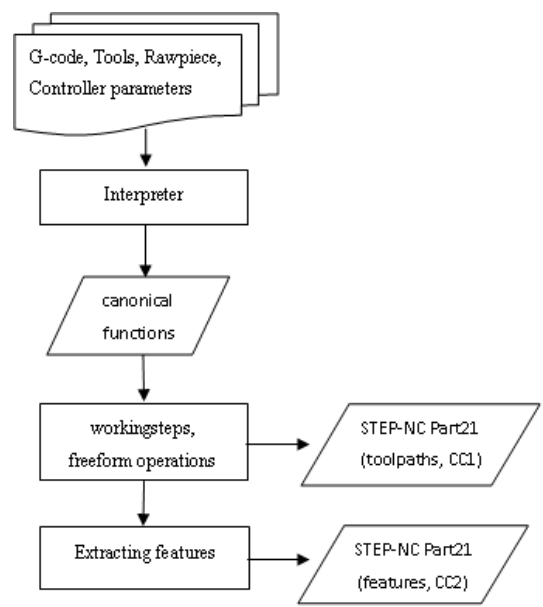

**Figure 5: The overall strategy.**

## 3- G-code to canonical functions

### 3.1 – Canonical functions

Canonical machining functions were defined by NIST with two objectives. First, all the functionality of common 3 to 5-axis machining centers had to be covered by them; for any function a machining center can perform, there has to be a way to tell it to do that function. Second, it should be possible to interpret an ISO 6983/RS274 compatible NC program into canonical function calls.

Canonical functions are atomic commands. Each function produces a single tool motion or a single logical action. NC commands include two types: those for which a single NC command corresponds exactly to a canonical function call, and those for which a single one will be decomposed into several canonical function calls. For instance, G1 (move in a straight line), M8 (turn flood coolant on) are of the first type; G83 (peck drilling) is of the second type.

Some main canonical functions are listed below:

*1) Representation: SET_ORIGIN_OFFSETS(); USE_LENGTH_UNITS()*
*2) Free Space Motion: STRAIGHT_TRAVERSE()*
*3) Machining Attributes: SELECT_PLANE(); SET_FEED_RATE()*
*4) Machining Functions: STRAIGHT_FEED(); ARC_FEED()*
*5) Spindle Functions: SET_SPINDLE_SPEED();*
*START_SPINDLE_CLOCKWISE();*
*START_SPINDLE_COUNTERCLOCKWISE();*
*6) Tool Functions: CHANGE_TOOL(); SELECT_TOOL();*
*USE_TOOL_LENGTH_OFFSET()*
*7) Miscellaneous Functions: FLOOD_OFF(); FLOOD_ON()*
*8) Program Functions: PROGRAM_STOP()*

In G-code, canned cycles (G81-G89) are for hole machining operations. A canned cycle is decomposed into its basic moves expressed by the above canonical functions. For example, (G81 X…Y…Z…R…L…) is intended for drilling. Its motions are: preliminary motion(s); move Z-axis at current feedrate to the Z position; Retract Z-axis at traverse rate to clear Z. This work does not handle G-codes programs that have macro-commands. Non-linear G-code program structure, like parallel, selective, non-sequential, are not handled, either.

### 3.2 – The role of the conversion

In a G-code program, the working coordinate system may change frequently, and the coordinate data may be absolute or relative, and the radius/length compensation may be used. All these are up to the user's choice according to programming convenience. In addition, as in all dialects of G-codes, a line of code may specify several different things to do, such as moving from one place to another along a line/arc, changing the feed rate, starting the spindle turning, etc. The Interpreter reads lines of a G-code one at a time, emulates the execution, and keeps track of the current state of the controller. We use the Interpreter: to check the correctness of the inputting G-code; to unify the coordinates and parameters for a line/arc move with respect to a chosen coordinate system; to determine the correct execution order of the G-code. The Interpreter outputs the following canonical functions as shown in Table 2, for the example G-code. This conversion facilitates greatly the subsequent processes.

**Table 2: Canonical functions for milling the planar face**

| | |
|---|---|
| 1 USE_LENGTH_UNITS(UNITS_MM) | 17 STRAIGHT_FEED(91.9, 133.5, 0) |
| 2 SET_ORIGIN_OFFSETS(0, 0, 0) | 18 STRAIGHT_FEED(74.8, 133.5, 0) |
| 3 SET_FEED_REFERENCE(CANON_XYZ) | 19 STRAIGHT_FEED(74.8, -13.5, 0) |
| 4 STOP_SPINDLE_TURNING() | 20 STRAIGHT_FEED(57.7, -13.5, 0) |
| 5 FLOOD_OFF() | 21 STRAIGHT_FEED(57.7, 133.5, 0) |
| 6 SPINDLE_RETRACT() | 22 STRAIGHT_FEED(40.6, 133.5, 0) |
| 7 USE_TOOL_LENGTH_OFFSET(0) | 23 STRAIGHT_FEED(40.6, -13.5, 0) |
| 8 CHANGE_TOOL(ENDMILL_18MM) | 24 STRAIGHT_FEED(23.5, -13.5, 0) |
| 9 USE_TOOL_LENGTH_OFFSET(50) | 25 STRAIGHT_FEED(23.5,133.5,0) |
| 10 FLOOD_ON() | 26 STRAIGHT_FEED(6.4, 133.5, 0) |
| 11 SET_SPINDLE_SPEED(720) | 27 STRAIGHT_FEED(6.4, -13.5, 0) |
| 12 START_SPINDLE_CLOCKWISE() | 28 STRAIGHT_TRAVERSE(6.4,-13.5,15) |
| 13 STRAIGHT_TRAVERSE(91.9,-13.5,100) | 29 STOP_SPINDLE_TURNING() |
| 14 STRAIGHT_TRAVERSE(91.9,-13.5,15) | 30 FLOOD_OFF() |
| 15 SET_FEED_RATE(240) | 31 SPINDLE_RETRACT() |
| 16 STRAIGHT_FEED(91.9, -13.5, 0) | 32 PROGRAM_END() |

## 4- Canonical functions to STEP-NC Part21

A STEP Part 21 file contains two sections: the header section and the data section. Since the conversion is only relevant to the data section, the following will only consider the data section of a Part21 file. The data section has the following main types of entities: cutting tools; definitions and their setups of rawpieces (usually only one); one project, workplans and workingsteps; manufacturing features; machining operations; technology, functions and machining strategies; and placements, planes, dimensions, etc. The entities for tools and rawpieces can be written in the Part21 file directly as per the inputting information since they have less connection involved with other entities.

We devised a canonical function interpreter for the generation of other entities. In the beginning a "struct" buffer is defined to keep track of the controller status, such as the units, the security plane, the coolant switch, feed rate, spindle speed, the current tool and its position. The interpreter maintains this buffer when it runs. The interpreter emulates the execution of the canonical functions one by one: if it meets a "STRAIGHT_TRAVERSE", a "rapid movement" entity is created; if it meets one or several consecutive "STRAIGHT_FEED"s or "ARC_FEED"s, a "machining workingstep" which includes a freeform operation is created. The parameters of these functions are used as cutter location data stored in the toolpath list of the freeform operation. Of course sometimes computation is needed for obtaining the toolpaths. Other data, such as technology and machining functions can be inferred from the buffer. The entities of "toolpath features" here are used for information only. For cycle operations, hole features, not toolpath features are used. Thus when the emulation comes to an end, we can get a STEP-NC Part21 CC1 file in explicit toolpaths (cf. Appendix A.1, here only the data section is shown).

## 5- STEP-NC Part21 file to manufacturing features

This phase deals with mainly the extraction of manufacturing/machining features from toolpaths based on the previous Part21 CC1 file. Yan [YY1] adopts a Z-map based method; Other type of method is: first build a CAD model from a simulated or cut model usually in a STL file, then using conventional methods do the recognition [**AY1**, SP1]. Here we carry out the work based on toolpaths and tool's geometry. In this phase, one freeform operation corresponds to one machining workingstep. Many freeform operations might correspond to one manufacturing feature because often there are several layers of a rough machining and finish machining, which are needed to make a final feature. So one major procedure is to merge those freeform operations that machine the same feature, as well as the relevant workingsteps and rapid movements.

The feature types we are coping with are the machining features defined in ISO 14649 Part 10[**I2**]. The procedure for processing phase 1 is detailed as follows.

Step 1: From a freeform operation, get the tool and the toolpath list. If no freeform operations left, end phase 1.

Step 2: Analyse the feature type by tool type first. If the tool type is for drilling, the feature is a round hole, and note down the diameter and the axis. They can be used to identify the same hole in other operations; if the tool is a facemill, the feature is a planar_face; if the tool is a T-slot_mill or dovetail_mill or woodruff_keyseat_mill, the feature is obviously a slot. Go to Step 1.

Step 3: If the tool is an endmill (including tapered_endmill, ball_endmill, bullnose_endmill), then analyse the x, y, z-values of the CL (cutter location) data in the toolpath list.

If z-value varies and x, y-values keep constant, it is milling a hole feature. Note down the diameter and axis. Go to Step 1.

Step 4: If x, y, z-values are all varying, it is a freeform milling operation to make a region (surface). Go to Step 6.

Step 5: If z-value keeps constant and x, y-values vary, it is a 2½D milling operation.

Step 6: Compare the toolpath list with the one of the next freeform operation, for each CL-data, if the corresponding x, y-values are the same and the z-value is decreased, the two operations are cutting different layers of the same feature, then they can be merged. For a region feature, go to Step 1.

After the above processing, extracted features are easily found; processing phase 2 deals with the remaining toolpaths to find planar_faces, general_outside_profiles, closed_pockets and open_pockets. A step feature is a special type of open_pocket that has only one wall face (the open boundary is a line segment). A slot feature is also a special type of pocket, whose profile shape has a constant width. The main grounds to find features in phase 2 are the cutting area (the tool's covering region for cutting movements, Figure 3b) and the milling strategy, which are computed based on toolpath CL data.

Common 2½D milling strategies are shown in Figure 6. Unidirectional milling is going from one side to the other, then lifting the tool and going back to the starting point. Bidirectional milling is in a zigzag fashion, i.e. going from one side to the other and back. Center milling is along the center of the feature. It is often used for milling along the center of a slot. Contour milling, which is a typical strategy for pocket milling, is in several paths following the contour of the feature. Contour spiral milling is similar to contour parallel milling, with the exception, that in this case the milling path is a truly spiral path rather than concentric paths which are connected by an orthogonal movement. In general practices, the relationship between features and milling strategies is as follows.

"closed_pocket", "round_hole" ——contour parallel milling, contour spiral milling.
"open_pocket" —— bidirectional/unidirectional milling.
"planar_face" —— bidirectional/unidirectional milling, contour parallel milling (outside-in).
"step" —— bidirectional/unidirectional milling.
"slot" —— center milling.
"general_outside_profile" —— contour parallel milling.

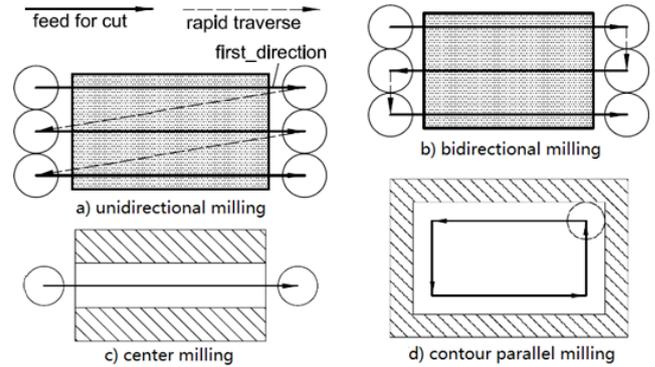

**Figure 6: Common milling strategies.**

In phase 2 the method is illustrated as in Figure 7. In 2½D manufacturing, the computation of the cutting area and milling strategy from toolpath CL data is not difficult since they are 2D geometries. The milling strategy can also be used for checking the correctness of the extracted features. After characterizing the feature types, the profiles of features are obtained from the boundaries of the cutting areas, which are determined by toolpaths and cutters' geometry.

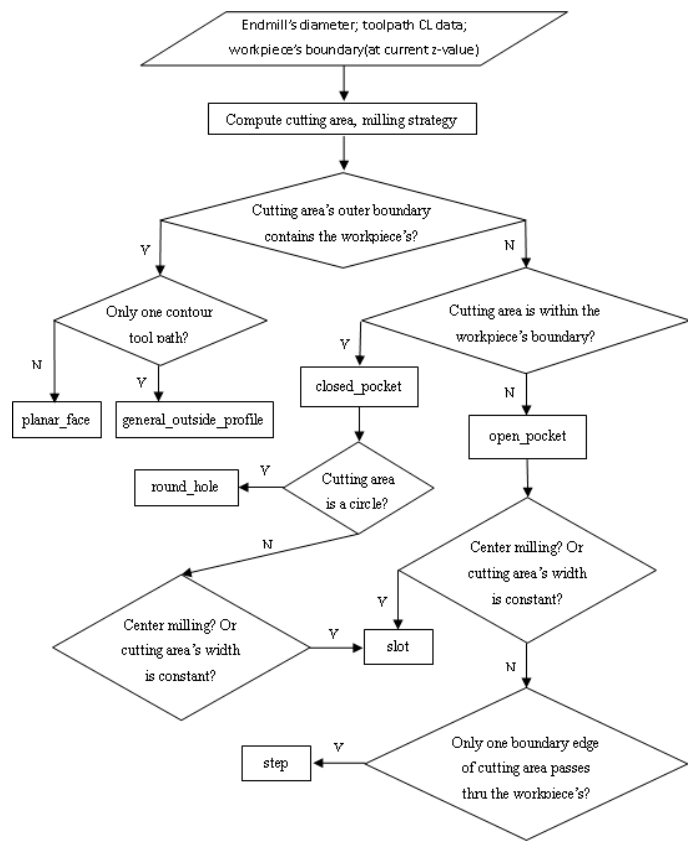

**Figure 7: Milled feature extraction.**

Sometimes the feature profiles thus obtained need to be fixed a bit if there is finish cutting or supplementary cutting to make the feature. For open features, such as planar_faces, steps, open_pockets, their profiles need not to be exact, as long as the open parts of their profiles are outside the workpiece's boundary. This requirement can simplify the computation. Some cutting parameters, as required by the machining technology in STEP-NC, such as radial/axial depths, can also be inferred from the milling strategy.

The converted STEP-NC Part21 CC2 file in manufacturing features is shown in appendix A.2 (only the data section shown). In this file the only feature is "PLANAR_FACE".

## 6- Implementation and examples

The entities of STEP-NC Parts 10, 11, 111 [I2] [I3] [I4] are mapped to C++ classes. The syntax of G-code and Part21 files is checked during the generation. The conversion of G-codes into canonical functions, and canonical functions into STEP-NC Part21 (CC1) can cope with 3 to 5-axis milling programs. The method for generating STEP-NC Part21 (CC2) is applied for 2½D G-code programs. In a future work, entities of STEP-NC Parts 12,121 (turning) will be mapped to C++ classes, so that turning programs can be also processed.

A test example is given in Figure 8. It has 3 machining features: a planar face, a round hole and a closed pocket. A section of the resulting Part 21 CC2 file is shown as follows, and part of the initial G-code program is shown in appendix A.3.

```
............
#10= MACHINING_WORKINGSTEP('WS FINISH PLANAR FACE1',#60,#20,#31,$);
#11= MACHINING_WORKINGSTEP('WS DRILL HOLE1',#60,#21,#32,$);
#12= MACHINING_WORKINGSTEP('WS REAM HOLE1',#60,#21,#33,$);
#13= MACHINING_WORKINGSTEP('WS ROUGH POCKET1',#60,#22,#34,$);
#14= MACHINING_WORKINGSTEP('WS FINISH POCKET1',#60,#22,#35,$);

#20= PLANAR_FACE('PLANAR FACE',#2,(#31),#64,#65,#23,#24,$,());
#21= ROUND_HOLE('HOLE1 D=22MM',#2,(#32,#33),#67,#70,#111,$,#25);
#22= CLOSED_POCKET('POCKET1',#2,(#34,#35),#69,#71,(),$,#26,$,#112,#27);
#23= LINEAR_PATH($,#110,#83);
#24= LINEAR_PROFILE($,#101);
#25= THROUGH_BOTTOM_CONDITION();
#26= PLANAR_POCKET_BOTTOM_CONDITION();
#27= RECTANGULAR_CLOSED_PROFILE($,#113,#114);

#31= PLANE_FINISH_MILLING($,$,'FINISH PLANAR FACE1',15.00,$,#40,#50,#51,$,
     #52,#52,#53,2.50,$);
#32= DRILLING($,$,'DRILL HOLE1',15.00,$,#44,#54,#51,$,$,$,$,#55);
#33= REAMING($,$,'REAM HOLE1',15.00,$,#47,#54,#51,$,$,$,$,#56,.T.,$,$);
#34= BOTTOM_AND_SIDE_ROUGH_MILLING($,$,'ROUGH POCKET1',15.00,$,#40,
     #57,#51,$,$,$,#58,6.50,5.00,1.00,0.50);
#35= BOTTOM_AND_SIDE_FINISH_MILLING($,$,'FINISH POCKET1',15.00,$,#40,#57,
     #51,$,$,$,#59,2.00,10.00,$,$);
............
```

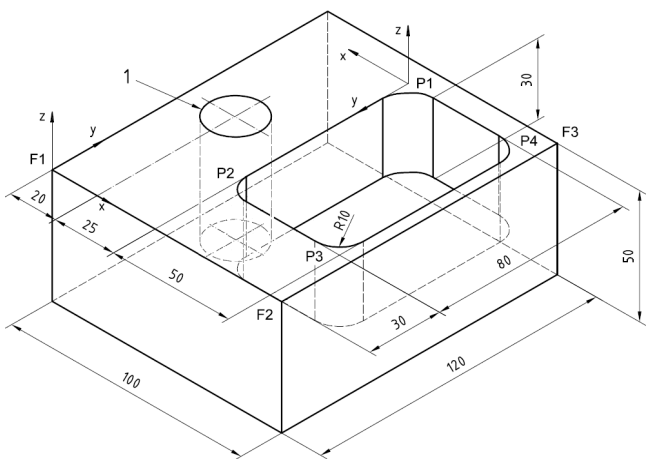

**Figure 8: A test example.**

## 7- Conclusions

While STEP-NC is gaining popularity in the manufacturing industry, the issue of legacy G-code program conversion is arising. This paper proposes an overall approach for the conversion of a G-code program into STEP-NC. The conditions for a valid conversion are detailed. This approach includes three main phases: first, G-code program is converted into canonical functions; then, canonical functions are analyzed to create a Part21 file in explicit toolpaths (CC1); and finally, by applying machining feature recognition techniques, the Part21 file is rewritten into a higher level Part21 file in manufacturing features (CC2). Although, 2½D machining features can be recognized, and the machining feature recognition approach developed here have no feature interaction issues, the problem of merging freeform operations that make the same feature has to be solved. Future work will include: (a) detecting turning or milling machining type should be automatic, as well as the number of axis involved in the program, (b) the extension of the approach to handle turning/mill-turn G-code programs, (c) to increase the capacity of manufacturing feature recognition, especially for the case of region features.

## Acknowledgements

This work is a part of the ANGEL FUI project recently funded by the French Inter-ministerial Fund and endorsed by top French competiveness clusters (SYSTEMATIC PARIS REGION "Systems & ICT", VIAMECA "Advanced Manufacturing" and ASTECH "Aeronautics & Space").

## 8- References

# Appendix

## A.1 – Part21 file (CC1)

```
DATA;
#0= PROJECT('EXECUTE EXAMPLE1',#1,(#2),$,$,$);
#1= WORKPLAN('MAIN WORKPLAN',(#10, ,#13),$,#3,$);
#2= WORKPIECE('CUBOID WORKPIECE',$,0.01,$,$,$,(#91,#92,#93,#94));
#3= SETUP('SETUP1',#62,#60,(#4));
#4= WORKPIECE_SETUP(#2,#63,$,$,());
#10= MACHINING_WORKINGSTEP('WS FINISH PLANAR FACE1',#60,#20,#30,$);
#13= RAPID_MOVEMENT('rapid after milling plane', #60,#401,$);
#20= TOOLPATH_FEATURE('FACE1:LAYER1',#2,(#30),#63,#64);
#30= FREEFORM_OPERATION(#101,$,'FINISH FACE1 L1',15.00,$,#40,#50,#51,$,$,$,$);

#40= MILLING_CUTTING_TOOL('ENDMILL_18MM',#41,(#43),80.00,$,$);
#41= TAPERED_ENDMILL(#42,4,.RIGHT.,.,F.,$,$);
#42= MILLING_TOOL_DIMENSION(18.00, $,$, 29.0, 0.0, $,$);
#43= CUTTING_COMPONENT(80.00,$,$,$,$);
#50= MILLING_TECHNOLOGY(0.04,.TCP.,$,-12.00,$,.F.,.F.,$);
#51= MILLING_MACHINE_FUNCTIONS(.T.,$,$,.F.,$,(),.T.,$,$,());

#60= PLANE('SECURITY PLANE',#61);
#61= AXIS2_PLACEMENT_3D('PLANE1',#90,#81,#82);
#62= AXIS2_PLACEMENT_3D('SETUP1',#80,#81,#82);
#63= AXIS2_PLACEMENT_3D('CUBOID WORKPIECE',#80,#81,#82);
#64= PLANE('PLANAR FACE1:DEPTH PLANE',#65);
#65= AXIS2_PLACEMENT_3D('PLANAR FACE1',#95,#81,#82);
#80= CARTESIAN_POINT('ORIGIN',(0.00,0.00,0.00));
#81= DIRECTION('K-VECTOR',(0.00,0.00,1.00));
#82= DIRECTION('I-VECTOR',(1.00,0.00,0.00));
#90= CARTESIAN_POINT('SECURITY PLANE:LOCATION',(0.00,0.00,100.00));
#91= CARTESIAN_POINT('CLAMPING_P1',(0.00,20.00,25.00));
#92= CARTESIAN_POINT('CLAMPING_P2',(100.00,20.00,25.00));
#93= CARTESIAN_POINT('CLAMPING_P3',(0.00,100.00,25.00));
#94= CARTESIAN_POINT('CLAMPING_P4',(100.00,100.00,25.00));
#95= CARTESIAN_POINT('PLANAR FACE1:LOCATION ',(0.00,0.00,0.00));

#101= TOOLPATH_LIST((#102));
#102= CUTTER_LOCATION_TRAJECTORY(.T.,,.TRAJECTORY_PATH.,$,#50,$,$,#103,$,$);
#103= POLYLINE('1st cut of planar FACE1',(#110,#111,#112,#113,#114,#115,#116,
    #117,#118,#119,#120,#121,#122));
#110= CARTESIAN_POINT('',(91.90, -13.50, 15.00));
#111= CARTESIAN_POINT('',(91.90, -13.50, 0.00));
#112= CARTESIAN_POINT('',(91.90, 133.50, 0.00));
#113= CARTESIAN_POINT('',(74.80, 133.50, 0.00));
#114= CARTESIAN_POINT('',(74.80, -13.50, 0.00));
#115= CARTESIAN_POINT('',(57.70, -13.50, 0.00));
#116= CARTESIAN_POINT('',(57.70, 133.50, 0.00));
#117= CARTESIAN_POINT('',(40.60, 133.50, 0.00));
#118= CARTESIAN_POINT('',(40.60, -13.50, 0.00));
#119= CARTESIAN_POINT('',(23.50, -13.50, 0.00));
#120= CARTESIAN_POINT('',(23.50, 133.50, 0.00));
#121= CARTESIAN_POINT('',(6.40, 133.50, 0.00));
#122= CARTESIAN_POINT('',(6.40, -13.50, 0.00));

#401=TOOLPATH_LIST((#402));
#402= CUTTER_LOCATION_TRAJECTORY(.T.,,.TRAJECTORY_PATH.,$,$,$,$,#403,$,$);
#403= POLYLINE('rapid after milling plane',(#410,#411));
#410= CARTESIAN_POINT('',(6.40, -13.50, 0.00));
#411= CARTESIAN_POINT('',(6.40, -13.50, 15.00));
ENDSEC ;
```

## A.2 –Part21 file (CC2)

```
DATA;
#0= PROJECT('EXECUTE EXAMPLE1',#1,(#2),$,$,$);
#1= WORKPLAN('MAIN WORKPLAN',(#10),$,#3,$);
#2= WORKPIECE('CUBOID WORKPIECE',$,0.01,$,$,$,(#91,#92,#93,#94));
#3= SETUP('SETUP1',#62,#60,(#4));
#4= WORKPIECE_SETUP(#2,#63,$,$,());
#10= MACHINING_WORKINGSTEP('WS FINISH PLANAR FACE',#60,#20,#30,$);

#20= PLANAR_FACE('PLANAR FACE1',#2,(#30),#64,#65,#21,#22,$,());
#21= LINEAR_PATH($,#23,#83);
#22= LINEAR_PROFILE($,#25);
#23= TOLERANCED_LENGTH_MEASURE(120.00,#24);
#24= PLUS_MINUS_VALUE(0.30,0.30,3);
#25= NUMERIC_PARAMETER('PROFILE LENGTH',100.00,'MM');

#30= PLANE_FINISH_MILLING($, $, 'FINISH PLANAR FACE1', 15.00, $, #40, #50, #51, $,
    #52, #52, #53, 2.50, $);

#40= MILLING_CUTTING_TOOL('ENDMILL_18MM',#41,(#43),80.00,$,$);
#41= TAPERED_ENDMILL(#42,4,.RIGHT.,.,F.,$,$);
#42= MILLING_TOOL_DIMENSION(18.00, $,$, 29.0, 0.0, $,$);
#43= CUTTING_COMPONENT(80.00,$,$,$,$);

#50= MILLING_TECHNOLOGY(0.04,.TCP.,$,-12.00,$,.F.,.F.,$);
#51= MILLING_MACHINE_FUNCTIONS(.T.,$,$,.F.,$,(),.T.,$,$,());
#52= PLUNGE_TOOLAXIS($);
#53= BIDIRECTIONAL_MILLING(0.05,.T.,#83,.LEFT.,$);
```

#60= PLANE('SECURITY PLANE',#61);
#61= AXIS2_PLACEMENT_3D('PLANE1',#90,#81,#82);
#62= AXIS2_PLACEMENT_3D('SETUP1',#80,#81,#82);
#63= AXIS2_PLACEMENT_3D('CUBOID WORKPIECE',#80,#81,#82);
#64= AXIS2_PLACEMENT_3D('PLANAR FACE1',#95,#81,#82);
#65= PLANE('PLANAR FACE1-DEPTH PLANE',#66);
#66= AXIS2_PLACEMENT_3D('PLANAR FACE1',#96,#81,#82);
#80= CARTESIAN_POINT('ORIGIN',(0.00,0.00,0.00));
#81= DIRECTION('K-VECTOR',(0.00,0.00,1.00));
#82= DIRECTION('I-VECTOR',(1.00,0.00,0.00));
#83= DIRECTION('J-VECTOR',(0.00,1.00,0.00));
#90= CARTESIAN_POINT('SECURITY PLANE:LOCATION',(0.00,0.00,100.00));
#91= CARTESIAN_POINT('CLAMPING_P1',(0.00,20.00,25.00));
#92= CARTESIAN_POINT('CLAMPING_P2',(100.00,20.00,25.00));
#93= CARTESIAN_POINT('CLAMPING_P3',(0.00,100.00,25.00));
#94= CARTESIAN_POINT('CLAMPING_P4',(100.00,100.00,25.00));
#95= CARTESIAN_POINT('PLANAR FACE1:LOCATION ',(0.00,0.00,5.00));
#96= CARTESIAN_POINT('PLANAR FACE1:DEPTH ',(0.00,0.00,-5.00));
ENDSEC;

## A.3 –Part of the initial G-code for the test

| | |
|---|---|
| G54 G90 G21 G40 G49 M5 M9 | (To rough pocket in 5 layers, 5.9/layer) |
| T1 M6 (Use an endmill, diameter 18mm) | (First 2 blocks: to run helical approach) |
| G43 H1 (Length compensation by 50mm) | G2 X77.2 Y55. Z-5.9 I5.246 J4.932 |
| M8 S720 M3 | G2 X70. Y55. I-3.60 J0. |
| G0 X91.9 Y-13.5 Z100. | G1 Y90. |
| | X75. |
| (To finish top face of rawpiece in 2 layers) | Y50. |
| Z15. | X65. |
| G1 Z2.5 F240. (1st layer, depth 2.5mm) | Y90. |
| Y133.5 | X70. |
| X74.8 | Y95. |
| Y-13.5 | X80. |
| X57.7 | Y45. |
| Y133.5 | X60. |
| X40.6 | Y95. |
| Y-13.5 | X70. |
| X23.5 | Y100. |
| Y133.5 | X85. |
| X6.4 | Y40. |
| Y-13.5 | X55. |
| G0 Z15. | Y100. |
| X91.9 | X70. |
| G1 Z0. (2nd layer, depth 2.5mm) | Z0. |
| Y133.5 | G0 X69.532 Y47.815 (End of 1st layer) |
| X74.8 | ………… (Code of next 4 layers omitted) |
| Y-13.5 | |
| X57.7 | (To finish pocket in 6 layers. 5mm/layer) |
| Y133.5 | (Bottom allowance 0.5, side allowance 1) |
| X40.6 | G0 Z30. |
| Y-13.5 | X74.890 Y60.285 |
| X23.5 | Z15. |
| Y133.5 | (First 2 blocks: to run helical approach) |
| X6.4 | G2 X77.2 Y55. Z-2. I-4.891 J-5.285 |
| Y-13.5 | G2 X70. Y55. I-3.60 J0. |
| | G1 Y93. |
| (To drill and ream a thru hole) | X78. |
| G0 Z15. | Y47. |
| G49 M9 M5 | X62. |
| T2 M6 (Use a spiral drill, diameter 20mm) | Y93. |
| G43 H2 (Length compensation by 70mm) | X70. |
| M8 M3 F900. S720 | Y101. |
| G0 Z30. | X85. |
| G90 G99 G81 X20. Y60. Z-18. R10. | G2 X86. Y100. I0. J-1. |
| G99 G81 X20. Y60. Z-36. R10. F1800. | G1 Y40. |
| G99 G81 X20. Y60. Z-60. R10. F1350. | G2 X85. Y39. I-1. J0. |
| G1 Z10. F1800. | G1 X55. |
| G80 G49 M5 M9 (end of drilling cycle) | G2 X54. Y40. I0. J1. |
| T3 M6 (Use a reamer, diameter 22mm) | G1 Y100. |
| G43 H3 (Length compensation by 50mm) | G2 X55. Y101. I1. J0. |
| M8 M3 S1080 | G1 X70. |
| G90 G99 G85 X20. Y60. Z-60. R10. | Z0. (End of 1st layer) |
| G80 G49 M5 M9 (End of reaming cycle) | ............(Code of rest of layers omitted) |
| | G2 X55. Y101. I1. J0. (now Z-30.00) |
| (To cut a pocket, rough & finish) | G1 Y70. |
| T1 M6 (Use an endmill, diameter 18mm) | Z15. (End of finishing) |
| G43 H1 (Length compensation by 50mm) | M30 |
| M8 S1200 M3 F2400. | |
| G0 Z30. | |
| X64.754 Y50.069 | |
| Z15. | |